\documentclass[twocolumn,showpacs,preprintnumbers,amsmath,amssymb,prb,aps]
{revtex4}
\pdfoutput=1
\usepackage{epsfig} %Include figure filesm 
\usepackage{graphicx}% Include figure files
\usepackage{dcolumn}% Align table columns on decimal point
\usepackage{bm}% bold math
\usepackage{epstopdf}
\usepackage{amsmath}
\usepackage{xcolor}
\begin{document}

\title{ Coupling of magnetism and transport properties to the lattice degrees of freedom in NdBaCo$_2$O$_{5+\delta}$ ($\delta \sim 0.65$)}
\author{Himanshu Pant$^1$}
\author{Saurabh Singh$^{2,3}$}
\author{Jaskirat Brar$^1$}
\author{Priyamedha Sharma$^1$}
\author{M. Bharath$^1$}
\author{Kentaro Kuga$^2$}
\author{Tsunehiro Takeuchi$^{2,3}$}
\author{R. Bindu$^1$}
\altaffiliation{Corresponding author: bindu@iitmandi.ac.in}
\affiliation{$^1$School of Physical Sciences, Indian Institute of Technology Mandi, Kamand, Himachal Pradesh- 175005, India}
\affiliation{$^2$Toyota Technological Institute, Nagoya, Aichi 468-8511, Japan}
\affiliation{$^3$Japan Science and Technology Agency, Kawaguchi, Saitama 332-0012, Japan}
\date{\today}

\begin{abstract}

We have studied the origin of zero volume expansion below the Curie temperature (T$_c$), variable range hopping (VRH) behaviour using structural, magnetic, transport and thermal studies on the oxygen deficient double perovskite NdBaCo$_2$O$_{5+\delta}$ ($\delta \sim$ 0.65). The valence state of Co ions and the possible properties exhibited by such compound were studied using electronic structure calculations for $\delta$ = 0.75. Careful investigation of structure shows that the compound stabilizes in tetragonal structure (\textit{P4/mmm}) having $2a_p \times 2a_p \times 2a_p$ (222) superstructure, where a$_p$ is the cubic perovskite lattice parameter. The compound exhibits a minimum in resistivity, ferromagnetic and ferrimagnetic transitions around 375 K, 120 K (T$_c$) and 60 K, respectively with signature of Griffiths phase above T$_c$. Our detailed structural analysis suggests signature of the onset of the above magnetic transitions at temperatures well above its stabilisation at long range level thereby leading to VRH behaviour. The observed zero thermal expansion in volume below T$_c$ appears to be due to competing magnetic interactions within and between the magnetic sublattices. Our electronic structure calculations show (a) the importance of electron - electron correlation in Nd 4\textit{f} and Co 3\textit{d} states (b) Co ions stabilize in intermediate spin (IS) state, having oxidation state less than +3 (c) half metallicity. Our results show the possibility of coupling between magnetism and ferroelectricity. We believe that our results especially on the valence state of the Co ion, zero thermal expansion in volume, short range magnetic orderings and the connection between different degrees of freedom will be helpful in clearing the ambiguities existing in literature on the nature of magnetism and thereby aiding in designing new functionalities.

\end{abstract}

\maketitle

\section{Introduction}

Double perovskites of the type A$_2$BB$’$O$_6$, AA$’$B$_2$O$_6$, AA$’$BB$’$O$_6$ have similar topology as the cubic perovskites but with different kinds of B/B$’$ and A/A$’$ ordering. Here, A, A$’$ refer to rare earth or alkaline earth ions; B, B$’$ mainly to the transition metal ions; and O, the oxygen ions. Depending on the nature of ordering, these kinds of compounds display fascinating properties like
metal to insulator transition (MIT)\citep{maignan1999, taskinreviewGBCO2005,yamada2022physical}, 
half metallicity\citep{Half_metal_BiCu3Mn4O12,Halfmetallic_Sr2CrReO6}, ferroelectricity\cite{FE_1,FE_2, FE_3}, 
ferromagnetism, 
good ionic conductivity\cite{Ionic_conductivity_ReBaco2O6-}, magnetoresistance\cite{MR_LaCu3Mn4O12,MR_Ca_LaCu3Mn4O12_DFT,maignan1999, taskinreviewGBCO2005}  
magneto dielectric effect\cite{Magnetodielectric_Re2NiMnO6}, 
magneto caloric effect\cite{Magnetocaloric_Re2BB'O6} etc. In the case of A$_2$BB$’$O$_6$, depending on the difference between the oxidation states of B and B$’$, the compound stabilizes in ordered, disordered or partially disordered arrangements\cite{lufaso2006structure,howard2003ordered}. The ordered arrangements can take simple patterns like rock salt, columnar or layered ordering.

In double perovskite compounds\cite{king2010_cation_ordering}, the A site ordering can be induced by octahedral tilting, anion vacancies, A site vacancies etc. The compounds that belong to RBaB$_2$O$_{5+\delta}$ stabilise in layered ordering of the rare earth R$^{3+}$ that is induced by anion site ordering. This depends on the ionic size mismatch of Ba$^{2+}$ and R$^{3+}$ ions. The compound with fully oxygen deficient R$^{3+}$ layer ($\delta$= 0) has the B cation in the square pyramidal environment and the coordination number of the R$^{3+}$ ion is reduced from 12 to 8. When $\delta$ = 0.5, there occurs equal amount of B cations with square pyramidal and octahedral coordination with the ordering of the anion vacancies of columnar type. Such kind of ordering has been observed in RBaCo$_2$O$_{5.5}$ compounds. The interesting aspect of the Co ion in the B site is that it exhibits spin state transitions with temperature in addition to the properties exhibited by rest of the transition metal ions. It has been observed that the size of the R ions dictates the amount of the additional oxygen that can be incorporated into the system. The ionic radii of Nd$^{3+}$ is such that it can have $\delta$ value as high as 1\cite{pralong2006_NBCO6}. For higher ionic radii, for example for La$^{3+}$, the cation ordering is lost. Hence, NdBaCo$_2$O$_{5+\delta}$ compound is favourable to study the properties related to cation and anion ordering as a function of $\delta$. The Nd based cobaltite that has $\delta$ = 0 and 0.5 requires special condition for its preparation, while the compound with $\delta$ higher than 0.5 can be prepared in air. Hence, we have chosen the NdBaCo$_2$O$_{5+\delta}$ compound that can be easily prepared, to study its properties.

 In the dc susceptibility measurements on NdBaCo$_2$O$_{5+\delta}$ ($\delta$ = 0.65,0.72)\citep{solin2022,lobanovsky2006}, a rise in the susceptibility has been observed $\sim$ 120 K and a peak is observed $\sim$ 100 K. In the literature, depending on the value of $\delta$, it has been reported that NdBaCo$_2$O$_{5+\delta}$ stabilizes in different crystal structures \citep{burley2003,lobanovsky2006}. Here, we discuss the neutron diffraction (ND) results reported earlier for $\delta \sim$ 0.69 and 0.75 compounds. For $\delta$ = 0.69, Burley \textit{et. al.}\citep{burley2003} have got the better representation for their neutron diffraction pattern using orthorhombic (O) structure having Pmmm space group, they have observed ordering along the c direction and signature of additional ordering along the b direction. It is also important to note that they could not satisfactorily explain negative thermal expansion of \textit{c}-parameter and the values of the magnetic moment. Based on the low temperature neutron diffraction results, Burley \textit{et. al.}\citep{burley2003} have shown G-type antiferromagnetism (AFM) for T $\leq$ 125 K.

Lobanovsky \textit{et. al.}\cite{lobanovsky2006}, have analysed the temperature dependent ND experiments carried out for $\delta$ = 0.75 using two models: (i) Model A -- \textit{Pmmm} space group with O - crystal structure having 222 superstructure and (ii) Model B -- coexistence of O - crystal structure with \textit{Pmmm} space group having $\delta$ = 0.5 and tetragonal (T) crystal structure with \textit{P4/mmm} space group having $\delta$ = 1. They have pointed out that with regard to the fitting, the reliability factor obtained for the single phase model is better than the two phase one. In the single-phase model proposed by Lobanovsky \textit{et. al.}\citep{lobanovsky2006}, the low temperature magnetism was attributed to canted G-type AFM structure based on Model A.

For $\delta$ = 0.75, Khalyavin et al.\citep{khalyavin2008} have reported their compound to be stabilized in Model A. They have suggested the existence of two sublattices of Co and Nd ions that order at T$_c \sim$ 180 K and T$_N \sim$ 40 K The former corresponds to the FM coupling within the sublattice and the latter to AFM coupling between the sublattices. Based on the magnetisation and ND results, they have come to the conclusion that the FM ordering in the Co sublattice is not uniform and  small AFM clusters also exists below $T_c$. It is important to note that despite Nd ions being magnetic, Burley \textit{et. al.}\citep{burley2003} and Lobanovsky \textit{et. al.}\cite{lobanovsky2006} in their low temperature ND studies have not mentioned about the magnetic ordering of the Nd sublattice. 

Zhang \textit{et. al.}\citep{zhang2023zero} studied the zero thermal expansion NdBaCo$_2$O$_{5.5+\delta}$ ($\delta$ = 0.18(3)) using neutron diffraction and magnetic measurements. Based on their results, they have reported coexistence of AFM and short-range FM interactions at low temperatures. However, it is important to note that the value of peak in the dc susceptibility data that corresponds to AFM transition is comparatively higher ($\sim$ 170 K) for the given $\delta$ value as compared to the ones reported in the literature\citep{burley2003,lobanovsky2006,khalyavin2008}. While Zhang \textit{et. al.}\citep{zhang2023zero} synthesized their sample using the sol-gel method, the other samples discussed above were prepared by the solid state route. This indicates that the method of synthesis plays an important role with regard to the magnetic properties as it is expected to induce inhomogeneities to different extents\cite{burley2003,lobanovsky2006,khalyavin2008,zhang2023zero,solin2022}.

The valence state of the compound also plays an important role in governing its physical properties. We now look into the valence state of the Co ion reported on NdBaCo$_2$O$_{5+\delta}$ compounds. For the case of $\delta$ = 0.5 compound, based on charge balance, all the Co ions are expected to stabilise in +3 oxidation state. For the case of $\delta<$ 0.5 and $\delta>$ 0.5, Co ions are expected to have additional mixed oxidation states of +2 and +4, respectively, along with the +3 state. It is interesting to note that, based on room temperature x-ray photoemission studies, Takubo \textit{et. al.}\citep{takubo2006} have reported that Co stabilizes only in +3 oxidation state in NdBaCo$_2$O$_{5+\delta}$ ($\delta$ = 0, 0.5). Hence, the oxidation state of Co ions in these double perovskite cobaltites is still an open question in both the hole ($\delta >$ 0.5 ) and electron ($\delta <$ 0.5) doped regions.

NdBaCo$_2$O$_{5+\delta}$ belong to the category of strongly correlated systems where there occurs significant interplay between the charge, spin, orbital and lattice degrees of freedom. As mentioned above, Khalyavin \textit{et. al.}\citep{khalyavin2008} in their single phase model have proposed ferrimagnetic (FeM) interaction between the Nd and Co sublattices. For the stabilisation of such interaction, variations in the associated structural parameters are expected. Also considering the covalent nature of the compounds, it is important to understand the valence states existing in such system thereby leading to the better understanding of its electronic and magnetic properties. Hence, we have prepared and performed temperature dependent x-ray diffraction, resistivity, heat capacity and magnetic measurements on NdBaCo$_2$O$_{5+\delta}$. To understand the valence state of Co in this compound and  behaviour of the states close to the Fermi level ($\varepsilon_f$), we have performed DFT+\textit{U} calculations.

Our results show that the compound crystallises in T-crystal structure (\textit{P4/mmm}) with 222 superstructure. We observe a clear link between structural, magnetic, transport and heat capacity. Our results also show the signature of short range magnetic ordering at temperatures well above the long range magnetic orderings. We also investigate the origin of negative thermal expansion in \textit{c}-parameter. Our electronic structure calculations show the significance of the electron-electron correlation in the Nd 4\textit{f} and Co 3\textit{d} states and signature of the stabilisation of the Co ion in less than +3 oxidation state and IS state. Signature of half-metallic behaviour is also observed in the calculations.

\section{Experimental}

Polycrystalline samples of  NdBaCo$_2$O$_{5+\delta}$ (of two different batches labelled as sample 1 and 2)  were prepared using the solid state reaction route in ambient atmospheric conditions. High purity powders of Nd$_2$O$_3$, BaCO$_3$ and Co$_3$O$_4$ ($>$ 99.99 \% purity purchased from Sigma-Aldrich)  were weighed in stoichiometric amounts and mixed. This mixture was hand ground for 5 hours using agate mortar and pestle, then pressed into pellets and sintered for 24 hours at 1000 $^o$C. Before weighing, Nd$_2$O$_3$, Co$_3$O$_4$ and BaCO$_3$ compounds were preheated at 850 $^o$C, 650 $^o$C and 400 $^o$C, respectively. The oxygen non-stoichiometry for sample 2 was obtained using iodometric titration\cite{maeno1991_Iodometric_titration}. It is important to note that the oxygen deficiency and inhomogeneities existing in the sample are highly dependent on the sample preparation conditions as highlighted by Burley \textit{et. al.}\cite{burley2003}

The x - ray diffraction (xrd) experiments were done using Smart Lab 9kW rotating anode x-ray diffractometer with Cu-K$_\alpha$ radiation (1.54 \AA). The low temperature xrd measurements were done inside a cryostat from room temperature (RT) to 10 K at 19 different temperatures for sample 1. For sample 2, xrd measurements were carried out at RT.

The Scanning Electron Microscopy (SEM) and Energy Dispersive X-ray Analysic (EDAX) measurements were performed on Nova Nano SEM-450 which is fitted with an energy dispersive X-ray spectrometer.

The temperature dependent dc susceptibility measurements were done using  Quantum Design SQUID, Magnetic Properties Measurement Systems (MPMS) in temperature range 300 K to 2 K at applied magnetic fields of 100 Oe and 1000 Oe for sample 1 and with Vibrating Sample Magnetometer (VSM) in temperature range 300 K to 2 K at applied magnetic fields of 100 Oe and 1000 Oe for sample 2. The field dependent magnetisation measurements up to 4 T were carried out using the Quantum Design SQUID
magnetometer for sample 1.

The temperature dependent resistivity measurements (using four probe method) and heat capacity measurements were carried out on sample 2 using Physical Properties Measurements System from Quantum design, USA.

\section{Computational Details}

Spin polarised density of states (DOS) calculations were performed using full potential linearised augmented plane wave basis with local-orbitals (LAPW + lo) as implemented in Elk code\cite{dewhurst2011_Elkcode}. For the exchange and correlation part, we have employed generalised gradient (GGA)- PBEsol approximation\cite{perdew2008_PBEsol}. On site Coulomb interaction term (\textit{U}) was used to take into account the strong electron-electron correlation in Nd 4\textit{f} and Co 3\textit{d} orbitals. The values of the $U_{Nd}$ were 2, 4, 6, 8 and 9 eV and $U_{Co}$ were 1, 2, 3 and 4 eV. Here $U_{Nd}$ and $U_{Co}$ mark the on site Coulomb interaction in Nd 4\textit{f} orbitals and Co 3\textit{d} orbitals, respectively. The experimental lattice parameters and the atomic positions were taken from our results of structural analysis for sample 1. The experimentally obtained atomic positions were relaxed initially in the spin polarised calculations with $U_{Nd}$ and $U_{Co}$ = 0 eV, keeping the lattice parameters fixed to the experimentally obtained values at RT, and these relaxed atomic positions were used for all the calculations. The muffin-tin sphere radii of  2.3, 2.3, 1.75 and 1.65 bohr were used for Nd, Ba, Co and O, respectively. For the termination of self consistent cycle, difference in  total energy was set to be less than 10$^{-4}$ Hartree/unit cell in the calculations. The structure was considered to be relaxed when the total force was less than 0.5$\times$10 $^{-3}$ Hartree/unit cell. These calculations were performed employing adaptive linear mixing using 5$\times$5$\times$5 k-point mesh for all the calculations.

\section{Results and discussion}

\subsection{Structural studies}

\begin{figure}[h!]

\includegraphics [width = 0.50\textwidth]{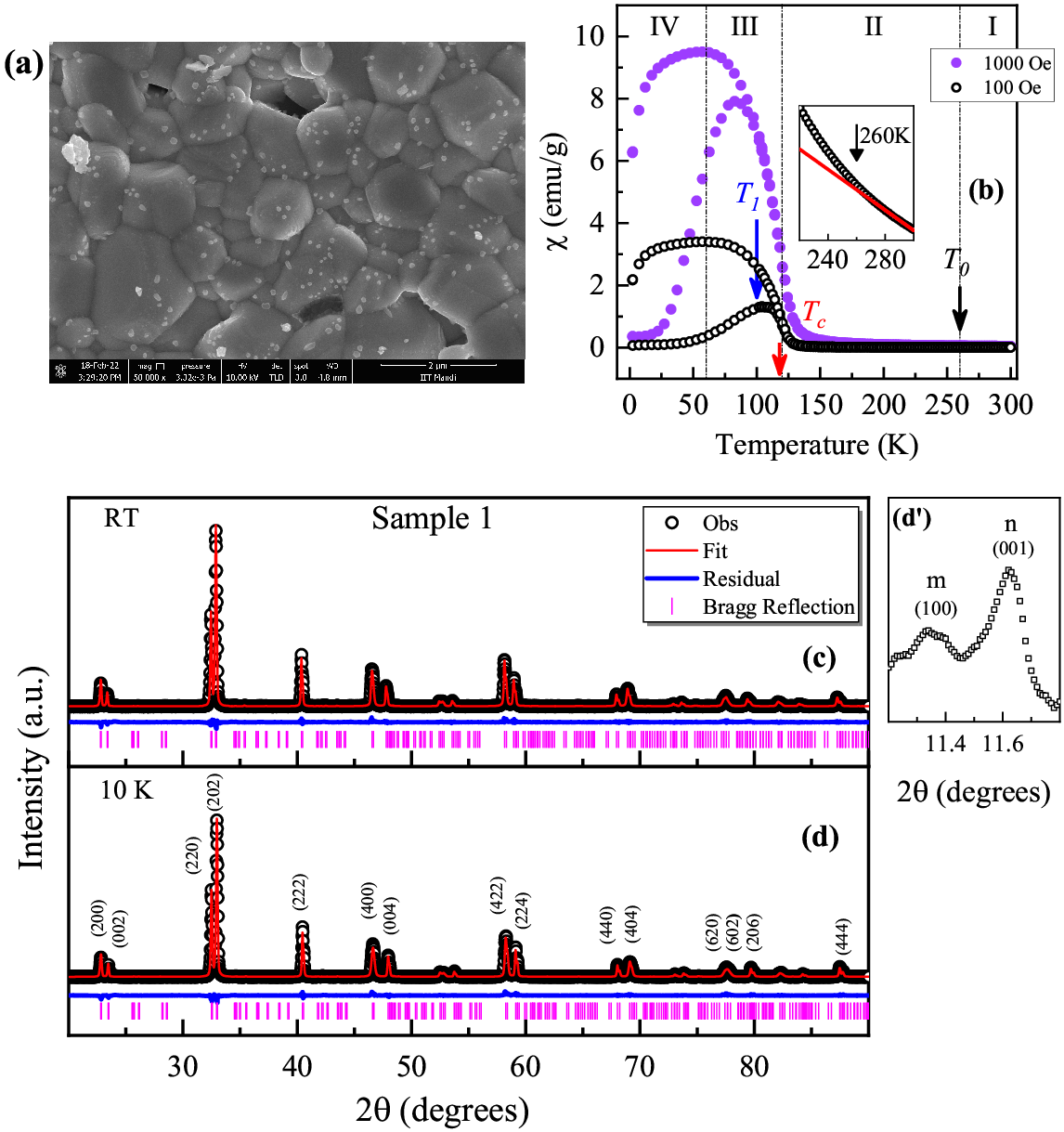}

\caption{For sample 1, (a) SEM image;
temperature dependent
(b) dc susceptibility at an applied magnetic field of 100 Oe and 1000 Oe during ZFC and FC; arrow marks $T_0$, $T_c$ and $T_1$; inset shows the plot in the temperature range 220 to 300 K for ZFC at 100 Oe.
Rietveld refinement of xrd patterns collected 
(c) at RT (300 K); 
(d) at 10 K; 
(d') shows the super lattice  peaks corresponding to oxygen vacancy ordering at RT. 
}

\end{figure}

In Fig.1 (a), the SEM image of the sample 1 is shown. The EDAX collected on different grains and on different regions of the sample shows that the ratio of Nd:Ba:Co was $\sim$ 1:1:2. The average (av.) grain size obtained is $\sim$ 1 µm.

The temperature dependent dc susceptibility measured during the zero field cooled (ZFC) and field cooled (FC) cycles at an applied field of 100 Oe and 1000 Oe for sample 1 is shown in Fig.1 (b). The behaviour of the susceptibility data is in line with the literature\citep{solin2022}. The $T_c$ is obtained from the first derivative of susceptibility with respect to temperature and the value obtained is $\sim$ 118 K at 100 Oe applied field. From the value of the $T_c$ obtained, the $\delta$ value based on literature\citep{solin2022} is estimated to be $\sim$ 0.65.

The RT xrd pattern of the compound (sample 1) is shown in Fig. 1(c). Based on the oxygen non-stoichiometry in the compound ($\delta \sim$ 0.65), it is expected to exhibit superlattice reflections in the 2$\theta$ range 10$^{\circ}$ to 12$^{\circ}$. To check for this possibility, RT (outside the cryostat) xrd experiments were carried out with slow scan (0.1$^{\circ}$/min with 0.01$^{\circ}$ step size) in this 2$\theta$ range. In this range we observe two peaks around 2$\theta$ = 11.4$^{\circ}$and 11.6$^{\circ}$ labelled m and n, respectively with peak n to be of higher intensity than peak m. The ratio of peak n to m obtained is $\sim$ 1.87. Similar ratios were observed on simulating the xrd patterns using the structural parameters previously reported for $\delta$ = 0.75\citep{khalyavin2008,frontera2004}, with O or T crystal structures having 222 super structure.
 
 In the case of O-crystal structure (\textit{Pmmm}), the superlattice reflections (010), (100) and (001) are expected to be observed around 11.34$^{\circ}$, 11.36$^{\circ}$ and 11.6$^{\circ}$, respectively. However, using conventional lab source xrd, the first two reflections of the 222 super structures are not expected to be resolved. Hence, only two peaks will be observed.

In the case of T-crystal structure (\textit{P4/mmm}) with 222 superstructure, the superlattice reflections (100) and (001) are expected to be observed around 11.4$^{\circ}$ and 11.6$^{\circ}$, respectively. It is important to note that this ratio calculated for O-structure (\textit{Pmmm}) reported for NdBaCo$_2$O$_{5.75}$ by Lobanovsky \textit{et. al.}\citep{lobanovsky2006} and  Khalyavin \textit{et. al.}\citep{khalyavin2008} comes out to be $\sim$ 3.58 and $\sim$ 6.66, respectively. In the xrd pattern calculated for PrBaCo$_2$O$_{5.75}$ having T-crystal structure (\textit{P4/mmm}) reported by Frontera \textit{et. al.}\citep{frontera2004}, the ratio comes out to be $\sim$ 3.81.

To identify the crystal structure for the compound under study, the RT xrd pattern was fitted using both the O and T-crystal structures with no oxygen vacancies having \textit{Pmmm} and \textit{P4/mmm} space groups, respectively and 222 superstructure. To obtain the $\delta$ value of 0.75 and identify the closest match to the experimentally obtained peak n to m ratio, the xrd patterns were calculated by removing oxygen from different sites. Our results show that for the case of T-crystal structure (\textit{P4/mmm}), with 222 super structure, by removing oxygen from 1b Wyckoff position, the value of the ratio obtained is closest to the experimental value i.e. $\sim$ 1.48. The structural parameters are given in Table-I.

The crystal structure of NdBaCo$_2$O$_{5.75}$ is shown in Fig. 2. In this structure, three kinds of Co ions are present and are labelled as Co1, Co2 and Co3. The Co1 ion stabilizes in square pyramidal environment and Co2 and Co3 are in octahedral environment with the oxygen ions. The Nd and Ba ions occupy alternate layers along the \textit{c}-axis.

In Fig.3(a-c),the temperature dependent lattice parameters, volume of the unit cell and \textit{c/a} ratio are shown. Based on the behaviour of \textit{c/a} ratio, temperature dependent lattice parameters are divided into 4 different regions. In region I (300 to 260 K), there occurs linear decrement of the \textit{c/a} ratio, in region II ($<$260 K to 120 K), there is non-linear and drastic decrement of \textit{c/a} ratio, in region III ($<$120 K to 60 K), an increment in this ratio is observed and in the region IV ($<$60 to 10 K) there is reduced increment in this ratio. With decrease in temperature, the lattice parameter \textit{a} shows a decrement until 260 K, a broad hump in the cross over region between region II and  III and in region IV, it remains almost the same. With temperature, lattice parameter \textit{c} shows a decrement in regions I and II and in region III, an increment is observed i.e. the \textit{c}-parameter shows a negative thermal expansion. The overall behaviour of the lattice parameters is qualitatively in line with neutron diffraction results reported by Burley et al.\cite{burley2003}. However, they could not explain the behaviour of the increment in the \textit{c}-parameter using the model that was chosen for fitting the diffraction pattern. It is also interesting to note that, there is almost zero thermal expansion in unit cell volume in the regions III and IV.

\begin{table}

\setlength{\tabcolsep}{1.5pt}
  \renewcommand{\arraystretch}{1.2}
  
  \caption{structural parameters obtained using Rietveld refinement for sample 1 and 2. Wyckoff positions - Ba 4j, Nd  4k, Co1 2g, Co2 2h, Co3 4i, O1 1a, O2 1c, O3 2f, O4 1b(unoccupied), O5 1d, O6 2e, O7 8s and O8 8t.  }
  
  \begin{tabular}{l|c|c|c}
  
  \hline 
  &\multicolumn{2}{c|}{\textbf{Sample 1}} & \textbf{Sample 2}\\
  &10 K & RT & RT\\
  \hline
  \textbf{Lattice parametrs} & & & \\
  
 \textit{ a=b} (\AA) & 7.7896(2)& 7.7989(2)& 7.8035(2)\\
  \textit{c} (\AA)& 7.5821(2) & 7.6081(2)& 7.6130(2)\\
  V (\AA$^3$)& 460.079(15) &  462.761(15)& 463.596(17)\\
  
  \textbf{Atomic positions} & & & \\
  
  Ba   (x)& 0.2509(3)& 0.2506(3) & 0.2440(5)\\
  Nd   (x)& 0.2565(3)& 0.2550(3)& 0.2499(4)\\
  Co1  (z)& 0.255(3)& 0.252(2)& 0.255(3)\\
  Co2  (z)& 0.254(3)& 0.254(2)& 0.253(3)\\
  Co3  (z)& 0.252(3)& 0.258(2)& 0.251(2)\\
  O7   (x)& 0.252(5)& 0.242(5)& 0.245(6)\\
       $\qquad$(z)& 0.285(3)& 0.286(3)& 0.277(4)\\
  O8    (x)& 0.243(5)& 0.252(5)& 0.239(5)\\
       $\qquad$(z)& 0.276(3)& 0.272(3)& 0.272(4)\\
       
  \hline
  
  \end{tabular}
   
\end{table}

\begin{figure}
\centering

\includegraphics [width = 0.45\textwidth]{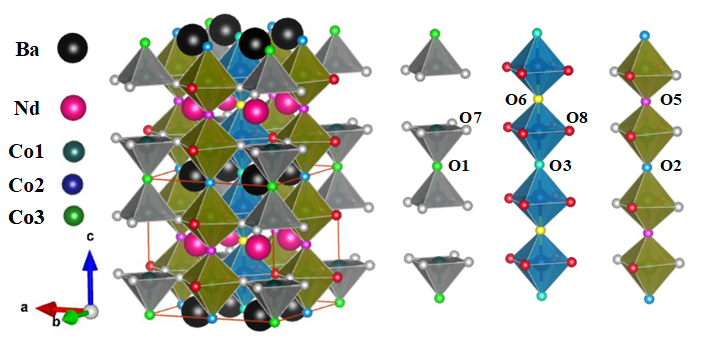}

\caption{Crystal structure of NdBaCo$_2$O$_{5.75}$.}

\end{figure}

\subsection{Magnetic studies}

 In Fig 1(b), temperature dependent dc susceptibility measured during the zero field cooled (ZFC) and field cooled (FC) cycles at an applied field of 100 Oe and 1000 Oe for sample 1 is shown. With decrease in temperature, in the FC cycle for both the applied fields, we observe a broad peak covering regions III and IV. The first transition temperature when magnetic moment starts to increase, we attributed this to the Curie temperature ($T_c$), transition from paramagnetic (PM) to ferromagnetic (FM) state. In addition to $T_c$, another broad peak is seen in the ZFC curve for 100 Oe applied field, $\sim$ 100K ($T_1$). This broad peak can be attributed to the onset of FM to ferrimagnetic (FeM) transition in the system. It is important to note that in the FC curve reported by Khalyavin \textit{et. al.}\citep{khalyavin2008}, have attributed the low temperature peak to such transition. Around the boundary of region III and IV (60K), at 100 Oe applied field, FC curve reaches a maximum in a broad temperature range and ZFC curve reaches a minimum. A deviation from PM behaviour at T$_0 \sim$ 260 K that falls in the boundary region of I and II is observed, shown in the inset of  Fig. 1(b). It is interesting to note that at the boundary of all the four regions, we observe slope change in the c/a ratio.

As both FM and FeM transitions are present in the system, a field induced transition is expected. Such a transition is reported in NdBaCo$_2$O$_{5.5}$\citep{solin2021NBCO55}. For confirming this, magnetic hysteresis was checked  at 5 K, Fig. 3(d). In this figure, a hysteresis, with coercive field of 0.77 T is observed. On careful observation during cycle 1 of this measurement, a change in slope is observed. This is clearly seen in inset of Fig. 3(d), where a peak is observed around 0.9 T in dM/dH vs H plot suggesting field induced transition from FeM to FM state i.e. metamagnetic transition.

\begin{figure}

\includegraphics [width = 0.50\textwidth]{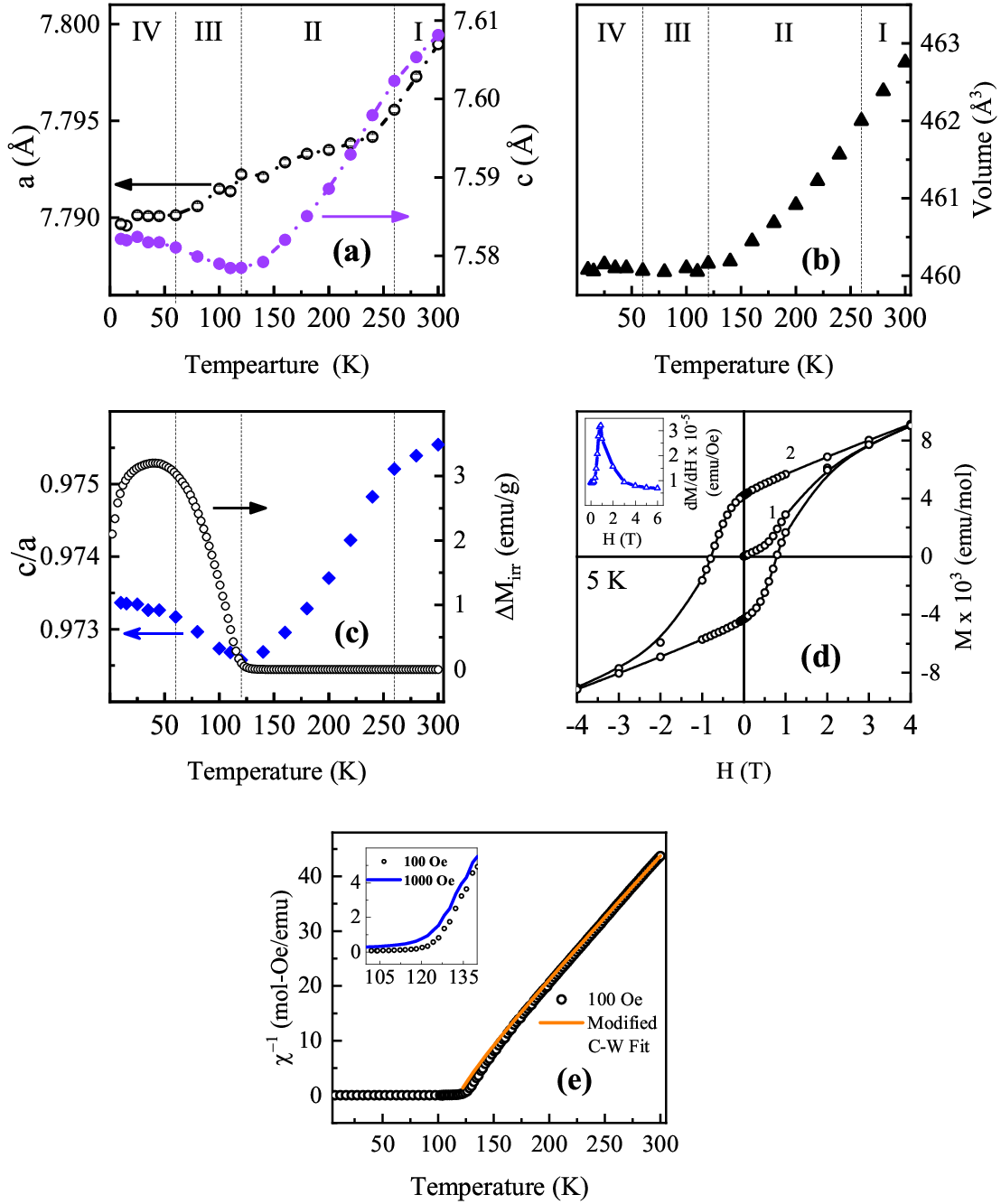}

\caption{ For sample 1, temperature dependent 
(a) lattice parameters; 
(b)unit cell volume; 
(c) tetragonal strain (c/a) and  $\Delta$M$_{irr}$. 
(d) magnetisation vs applied magnetic field at 5 K ; inset shows derivative of magnetisation with respect to field.
(e) inverse molar susceptibility at 100 Oe during FC and extrapolated modified C-W fit till $T_c$, inset shows the inverse molar susceptibility at an applied field of 100 and 1000 Oe in FC in 100 K to 140 K range. }

\end{figure}

In the compound under study, due to oxygen non-stoichiometry, one can expect local magnetic ordering. To examine this, temperature dependent inverse susceptibility is plotted in Fig. 3(e). In the figure, we observe a downturn from linear behaviour suggesting the stabilisation of short range ordering at temperatures much greater than $T_c$. In literature\cite{chakraborty2016disordered}, such short range ordering has been attributed to Griffiths phase.

Considering the fact that Nd and Co ions are magnetic in nature, the PM region, i.e., region I ($>$ 270 K)  was modelled using Curie- Weiss law given by 

\begin{equation*}
  \chi(T)= \dfrac{C{_{{Nd}}}}{T-\theta{_{{Nd}}}} + \dfrac{C{_{{Co}}}}{T-\theta{_{{Co}}}}
\end{equation*}

where C${_{{Nd}}}$ , C${_{{Co}}}$ are the Curie constants and $\theta{_{{Nd}}}$, $\theta{_{{Co}}}$ are the Curie temperatures for the Nd and Co ions, respectively. It is important to note that the C${_{{Co}}}$ corresponding to octahedral and square pyramidal environment were assumed to be the same. In the fitting, C${_{{Nd}}}$ and $\theta{_{{Co}}}$ was fixed to the known values, namely 1.62 and 118 K, respectively and  C${_{{Co}}}$ = 2.93(2) and $\theta{_{{Nd}}}$ = 60.07(2) K was obtained, table II. The latter temperature, possibly represents the FM ordering of the Nd$^{3+}$ ions but coupled ferrimagnetically with the spins of the Co ions. This observation is in line with the two sub lattice model proposed in neutron diffraction study by Khalyavin \textit{et. al.} for NdBaCo$_2$O$_{5.75}$\cite{khalyavin2008}. This explains the FeM behaviour observed below 60 K.

%To estimate the strength of Griffiths  phase, we have calculated the Griffiths exponent by using the modified Curie-Weiss law given by

%\begin{equation*}
%  \chi^{-1}= A{(T-T^{R}_C)}^{1-\lambda}
%\end{equation*}

%where $\lambda$ is Griffiths exponent, which lies in between 0 and 1 in Griffiths phase region and is 0 in PM region. The exponent $\lambda$ gives an estimate for deviation from the Curie-Weiss behaviour. T$^{R}_C$ is the critical temperature for random ferromagnet with no rare regions.\cite{kumar2020,bhoi2013,pramanik2010}. The value of $\lambda$ was obtained from Fig. 3(f), where T$^{R}_C \sim \theta_{Co}$. The value of $\lambda$ obtained from the linear fit in the PM phase is $\sim$ 0.07. In the Griffiths phase region the $\lambda$ obtained is $\sim$ 0.89.

\begin{table}

\setlength{\tabcolsep}{2 pt}
  \renewcommand{\arraystretch}{1.5}

  \caption{Parameters obtained after modelling the PM region using Curie-Weiss law.}
  \begin{tabular}{lcccc}
  
  \hline  
  &$T_c$($\theta_{Co}$) & $\theta_{Nd}$ &C$_{Co}$ &$\mu_{eff}^{exp} (Co)$ \\
  \hline
  \textbf{Sample 1}& 118 K & 60.07(2)&2.93(2)&4.84$\mu_B$\\
  \textbf{Sample 2}& 123 K & 60.19(2)&2.25(2)&4.24$\mu_B$\\
  \hline
  \end{tabular}
  
\end{table}

To further understand the physical properties of the compounds, additional measurements were carried out. It is important to note that to obtain single phase compound, only 1 gram could be prepared in a single batch. Using this quantity of sample, only temperature dependent xrd and magnetic studies were carried out. Hence, to perform transport studies, another batch of the same compound (sample 2) was prepared and characterised using iodometric titration, RT x-ray diffraction, dc susceptibility, resistivity and heat capacity measurements. 

Rietveld refinement of RT xrd pattern of the sample 2 is shown in Fig. S1(a). The SEM carried out on sample 2 also gives av. grain size $\sim$ 1 micron. The EDAX collected on different grains and in different regions of the sample also shows that the ratio of Nd:Ba:Co is $\sim$ 1:1:2 in the compound. This suggests that the sample is homogeneous and is of single phase. Iodometric titration for sample 2 gives $\delta \sim 0.64(2)$ which is near to the value estimated for sample 1 via dc susceptibility previously. The behaviour of the dc susceptibility is similar to that of sample 1 but with a small shift in the $T_c$ of $\sim$ +5 K and the decrease in value of the magnetic moment, shown in Fig. S2(a). Such change in the behaviour of $T_c$ can be due to the slight change in the oxygen non-stoichiometry\cite{solin2022}. The sample 2 also exhibits signature of Griffiths phase as shown in Fig S2(b). The results of the analysis of temperature dependent inverse susceptibility for samples 1 and 2 are given in Table II.

 Fig. S2(a), ZFC and FC curves show a bifurcation below $T_c$ in region III. To visualise this bifurcation, we have plotted temperature variation of the extent of bifurcation, $\Delta$M$_{irr}$ i.e. M$_{FC}$-M$_{ZFC}$ for different applied magnetic fields as shown in Fig. S2(c)\&(d). Our results show that with decrease in temperature, a sudden increase in the $\Delta$M$_{irr}$ is observed $\sim$ 132 K for 100 and 1000 Oe applied magnetic fields suggesting the presence of magnetic anisotropy in the system. Additionally, it was found that $\Delta$M$_{irr}$  increases with increasing applied field from 100 to 1000 Oe. However, it is expected to have no bifurcation for the applied fields higher than $\sim$ 0.77 T for the both samples as it marks the coercive field observed for sample 1. 

Our results show that an increase in $\Delta$M$_{irr}$ can be observed around $\sim$ 240 K for both the applied fields. This transition falls in region II, where Griffiths phase is expected to be dominant. It is interesting to note that when we combine Fig. 3(c) and S2(c-d), in the regions where anomaly in temperature dependent $\Delta$M$_{irr}$ is observed, we observe associated change in slope of the \textit{c/a} ratio. This suggests strong link between the tetragonal strain and magnetic anisotropy. To identify the nature of magnetic anisotropy, neutron diffraction experiments and ac susceptibility experiments will be helpful.

Despite the small changes in the values of $T_c$ and magnetic moment between samples 1 and 2 and also considering the fact that the nature of transition is same in both the samples, we expect that these results will not affect in the understanding of the link between the structural, magnetic and transport properties in this compound.

\subsection{Resistivity}

\begin{figure}

\includegraphics [width = 0.48\textwidth]{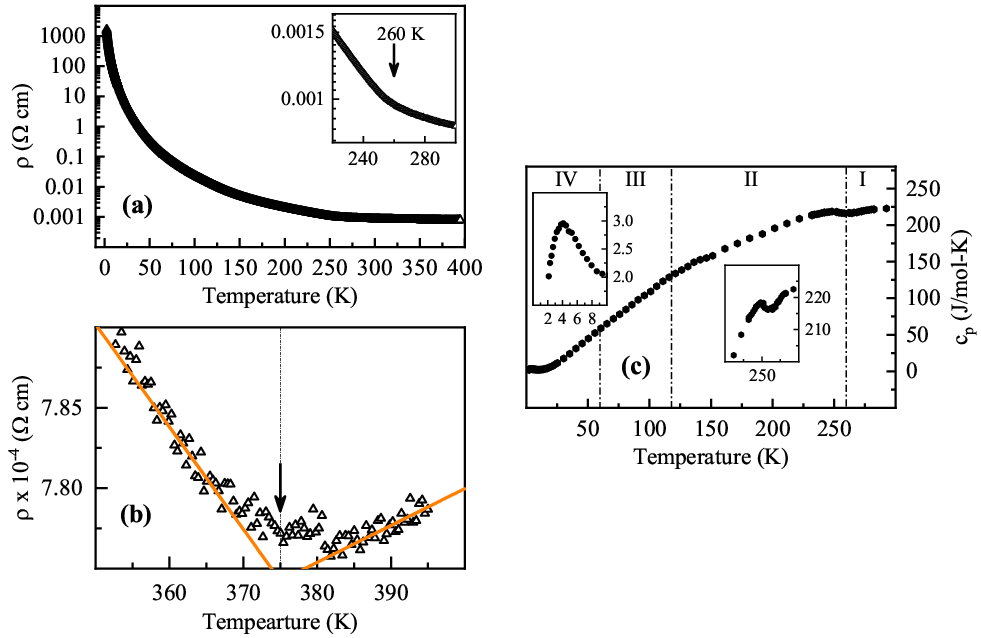}
\caption{For sample 2, temperature dependent  
(a) resistivity (warming cycle); inset shows the resistivity in the temperature range 220 to 300 K; 
(b) resistivity showing a minimum $\sim$ 375 K;
(c) heat capacity; inset shows the heat capacity in temperature range (left) 1-10 K, (right) 200-300 K.}

\end{figure}

The temperature dependent resistivity measurements were carried out on sample 2 in the temperature range 400 to 2.3 K. The sample exhibits a minimum in resistivity $\sim$ 375 K, Fig 4(a)\&(b). To understand the nature of this behaviour detailed temperature dependent resistivity and photoemission experiments are required. In the region below 375 K, the resistivity data was analysed using different models. 

In the temperature range 370 to 280 K, covering region I, the resistivity data was fitted using small polaron hopping model, Fig. 5(a). In amorphous and crystalline solids, when charge carriers self trap themselves in a potential created by their own interaction with the lattice then they are called polarons. And when the extent of the wave function of the charge carriers in space is less or comparable to the separation between the atoms, they are called small polarons\cite{emin1982small}. In disordered systems above $\theta_{D}$/2, the hopping due to small polarons is generally observed which follows the equation for resistivity as\cite{Polaronkhan2011,souza2008polaron}

\begin{equation*}
 \dfrac{\rho}{T} = \rho_{\alpha} \exp({{E_{p}}/{T}})
\end{equation*}

where , $\theta_{D}$ is Debye temperature and $E_p$ is activation energy.

$E_p$ = $W{_H}$ + {$W{_D}$}/2 (for $T > {\theta_{D}/2}$) where $W{_H}$ is polaron hopping energy and $W{_D}$ is disorder energy.

The prefactor $\rho_\alpha$ is defined as\cite{souza2008polaron} 

\begin{equation*}
 \rho_\alpha = \dfrac{{k_B}a}{{g_d}{e^2}{c(c-1)}{\nu_{ph}}}
\end{equation*}

where $k_B$ is Boltzmann constant, $g_d$ is geometrical factor, c is fraction of polarons and \textit{a} is hopping distance and $\nu_{ph}$ is optical phonon frequency given by  $\nu_{ph} \sim \dfrac{k_{B}\theta_{D}}{h}$.

We estimated $\theta_{D}$ from Fig 5(a), where starting value of linear fit in low temperature gives the value for $\theta_{D}$/2, which comes out to be $\sim$ 280 K. \citep{small_polaron_holstein1959, Polaronkhan2011} Hence $\nu_{ph}$ comes out to be 1.134 x 10$^{13}$ Hz.

In the system when polaron hopping probability is large and hopping takes palce every time when the nearest neighbour position gets the similar lattice distortion then hopping is said to be in the adiabatic limit. In the adiabatic limit the pre-factor ($\rho_\alpha$) is constant as defined above. While in non-adiabatic limit polaron hopping probability is small, pre-factor becomes temperature dependent and additional T$^{1/2}$ is added to it. To know the suitable limits in which the resistivity data is valid, Holstein condition\citep{small_polaron_holstein1959, Polaronkhan2011} is checked, according to which, polaron bandwidth J follows

\begin{align*}
 J>\phi, \quad & for~\text{Adiabatic small polaron hopping}		\\
 J<\phi, \quad & for~\text{Non-Adiabatic small polaron hopping}
\end{align*}

where \begin{equation*}
  \phi = \left(\dfrac{2k_B T W_H}{\pi}\right)^{1/4}\left(\dfrac{h\nu_{ph}}{\pi}\right)^{1/2}
\end{equation*}

 $W_H$ is polaron hopping energy. For small polaron formation \textit{J} and $W_H$ should satisfy the condition \textit{J} $<$ $W_H$/3. For the limiting case we have considered $W_{H}~ \sim $ 3\textit{J}.\citep{small_polaron_holstein1959, Polaronkhan2011}
 
 Temperature dependent polaron bandwidth is given by
 \begin{equation*}
  J(T) =0.67\ h \nu_{ph}\left(\dfrac{T}{\theta_D}\right)^{1/4}
\end{equation*}
By substituting the value \textit{J}(\textit{T} = 300 K) for the present case comes out to be $\sim$ 27 meV and value for  $\phi$ comes out to be $~\sim$ 23 meV. This satisfies the condition for adiabatic small polaron hopping.

\begin{figure}

\includegraphics [width = 0.48\textwidth]{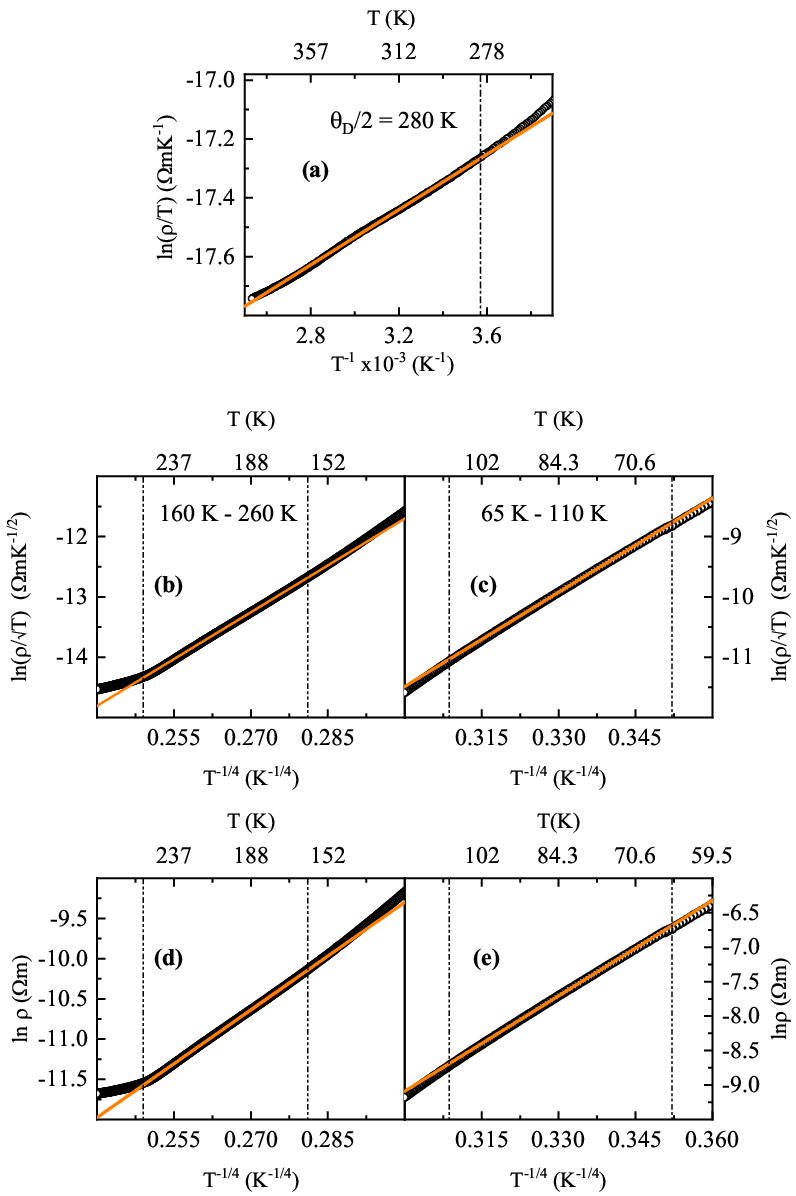}

\caption {Variation of resistivity for sample 2 showing (a) adiabatic small polaron hopping in temperature range 280 K - 375 K; Greaves variable range hopping conduction in temperature range (b) 260 K - 160 K; (c) 110 K - 65 K; Mott's 3D variable range hopping in temperature range (d) 260 K - 160 K; and (e) 110 K - 65 K.}

\end{figure}

As mentioned earlier, the compound under study shows Griffiths phase in region II, and also presence of oxygen vacancies in the system makes it a highly disordered system. To check for the contribution of disorder in this region, variable range hopping (VRH) models were utilised. We explored the possibility of Mott VRH model and Greaves VRH model in region II.

In the intermediate temperature range ($\theta{_D}/4$ $<$ T $<$ $\theta{_D}/2$) Greaves VRH model is given by

\begin{equation*}
 \dfrac{\rho}{\surd{T}} =\rho_{g} \exp{(T_g/T)^{1/4}}
\end{equation*}

 As shown in Fig. 5(b), Greaves VRH model is fitted in region II ($\sim$ 260 K - 160 K). This behaviour occurs because in lower temperatures lattice vibrations (phonons) which are helping in polaron hopping are reduced which in turn makes disorder energy to dominate over polaron hopping energy and VRH dominates.\cite{greaves1973small}
Interestingly, in region II, for the similar temperature range, data can also be fitted with 3D Mott VRH model, given by

\begin{equation*}
  \rho=\rho_{0} \exp{(T_0/T)^{1/4}}
\end{equation*}

where $\rho_{0}$ is constant pre factor and $T_0$ is characteristic temperature given by

\begin{equation*}
 T_0 = \dfrac{18}{k_BN(E_f)a^3}
\end{equation*}

Fig. 5(d) shows the graphs between ln$\rho$ and T$^{-1/4}$. From the fit $\rho_0$ and $T_0$ comes out to be 1.52$\times$10$^{-10}$ $\Omega$m and 3.86$\times$10$^{6}$ K, respectively in region II. For 3D VRH to be applicable, av. hopping distance ($R_M$) divided by localisation length, should be greater than unity. That is expressed as

\begin{equation*}
 \dfrac{R_M}{a} = \dfrac{3}{8}({T_0}/{T})^{1/4} > 1
\end{equation*}

This condition is checked in the limits of our fit, which validates VRH transport in the sample. It is interesting to note that both the Greaves and Mott VRH models can be applied to explain the transport in the system. However, such situation could be due to different nature of disorder present in the compound.

In region III, where sample shows a FM like transition (T${_c}\sim$ 123 K) we clearly observe a deviation from linear fit. In region II also the resistivity data follows both Greaves and Mott VRH  models, but with a change in the parameters as given in Table III, Fig. 5(c)\&(e). 

In region IV, the data could not be fit with any existing model in literature, possibly due to other effects coming into the picture.

\begin{table*}

\setlength{\tabcolsep}{1.5 pt}
  \renewcommand{\arraystretch}{1.2}
  
\caption{Results of the resistivity fitting.}

  \begin{tabular}{|c|c|c|}
  
  \hline 
  \textbf{Region I}&\textbf{Region II}&\textbf{Region III}\\
  375 K - 275 K & 260 K - 160 K & 110 K - 65 K\\
  (Adiabatic small polaron hopping)& (Variable range hopping) & (Variable range hopping) \\
  \hline
  &\textbf{Greaves}&\textbf{Greaves}\\
  E$_p$ $\sim$ 470 K &  T$_g$ $\sim$ 7.1 $\times$ 10$^{6}$ K & T$_g$ $\sim$ 7.3 $\times$ 10$^{6}$ K\\
  
  $\rho{_\alpha}$ $\sim$ 5.9 $\times$ 10$^{-9}$ $\Omega m K^{-1}$ & $\rho{_g}$ $\sim$ 1.5 $\times$ 10$^{-12}$ $\Omega m K^{-1/2}$& $\rho{_g}$ $\sim$ 1.7 $\times$ 10$^{-12}$ $\Omega m K^{-1/2}$\\
  
  $\theta{_D}$ $\sim$ 560 K &\textbf{Mott}&\textbf{Mott}\\
  & T$_0$ $\sim$ 3.9 $\times$ 10$^{6}$ K & T$_0$ $\sim$ 4.5 $\times$ 10$^{6}$ K \\
   
   & $\rho{_0}$ $\sim$ 10$^{-10}$ $\Omega m$& $\rho{_0}$ $\sim$ 10$^{-10}\Omega m$ \\
   
  \hline
  \end{tabular}

  \end{table*}

\subsection{Heat Capacity}

Temperature dependent heat capacity for sample 2 is shown in Fig. 4(c). To obtain the phonon contribution to heat capacity at high temperature, a combination of Debye and Einstein model is essential\cite{gutowska2016,wieckowski2012}. Considering the fact that there is a metal-insulator transition around $375\;$K and with the available data, it was not possible to perform such a fit. On careful observation, three peaks can be observed, (a)  $\sim$ 250 K (right inset of Fig. 4(c)), (b) a broad one around  $T_c$ ($\sim$ 123 K) and (c) $\sim$ 5 K (left inset of Fig. 4(c)). It is interesting to note that the first two peaks are associated with the transitions observed in the structural, magnetic and transport results, Fig. 1(b), 3(a) \& 5. The third peak observed around 5 K is associated with Schottky anomaly.\cite{wieckowski2012,gutowska2016} This is associated with thermal excitations of magnetic Nd$^{3+}$ ions to different levels of their ground state multiplet that splits under the influence of the crystalline field and Nd-Co exchange interactions\cite{gutowska2016,wieckowski2012}.

\begin{figure}
\centering

\includegraphics [width = 0.48\textwidth]{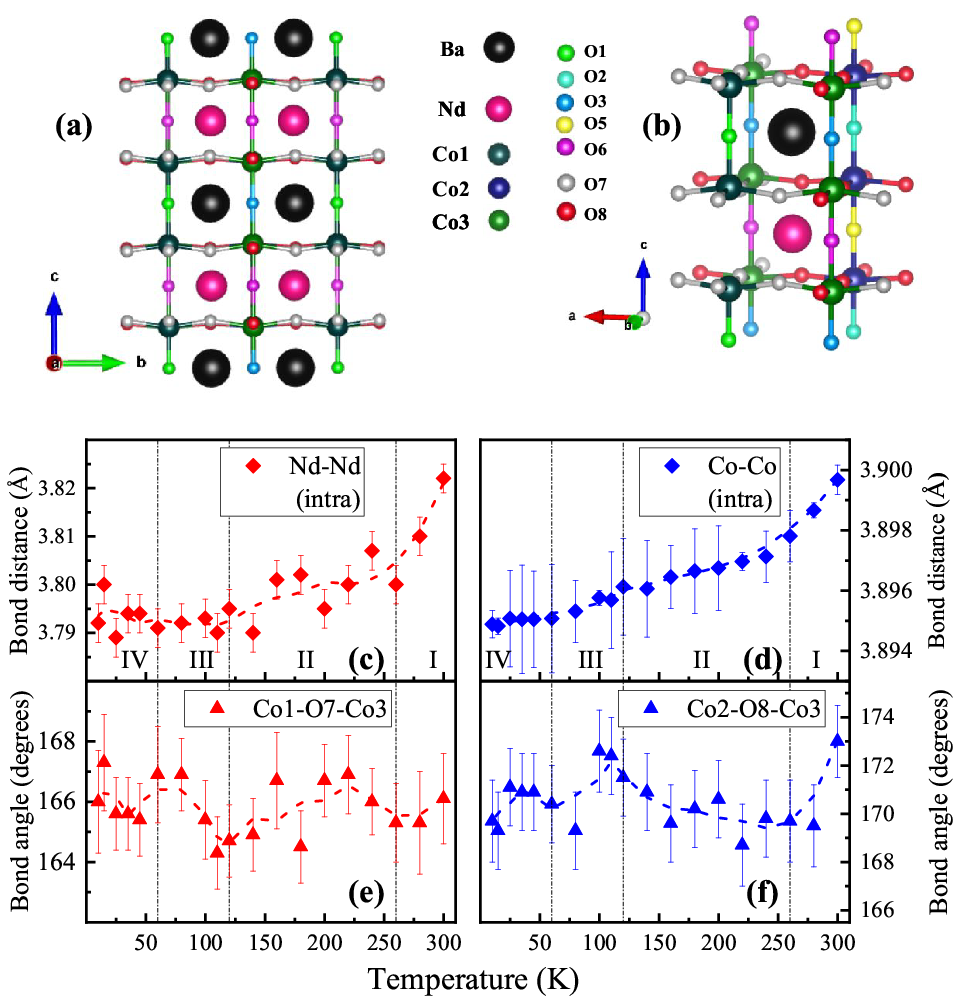}

\caption{ (a) The structure when viewed along the \textit{a}-axis. Here, different sizes of Nd and Ba cages can be clearly seen. (b) Structure showing one Ba and Nd cage only, for clarity. Temperature dependent (c) av. Nd - Nd (d) av. Co - Co distance in the ab plane. Temperature dependent basal bond angles (e) Co1-O8-Co3; (f) Co2-O7-Co3. }

\end{figure}

We now give a brief description of the crystal structure for a better understanding of its connection with the magnetism. In the crystal structure shown in Fig.6 (a), for the cage associated with the Nd$^{3+}$ or Ba$^{2+}$ ions, there are 6 Co ions in octahedral coordination and 2 Co ions with square pyramidal coordination. The Co1 and Co2, Co3 ions have square pyramidal and octahedral coordination with oxygen ions, respectively. Each of the Co3 ions is connected to two Co1 ions through the O7 ions and two Co2 ions through the O8 ions. The Ba$^{2+}$ and Nd$^{3+}$ ions are coordinated to 12 and 11 O ions, respectively. Owing to the bigger ionic radii of Ba$^{2+}$ ions, the cage size of the Ba$^{2+}$ ions is bigger than that of the Nd$^{3+}$ ions, Fig. 6(a-b). This increase in the size of the Ba$^{2+}$ cage as compared to the Nd$^{3+}$ cage is also manifested in the direction of the buckling of the Co-O-Co bond angles. In the case of the former, it is bent outward of the cage while in the case of the latter, it is the opposite, Fig. 6(a-b).
 
In region-I, with decrease in temperature, we observe decrement in both the lattice parameters due to thermal effect. Now, as the compound enters region-II, a broad maximum around $T_c$ is observed in the \textit{a}-parameter, while the \textit{c}-parameter shows a significant decrement but with a change in slope as compared to the decrement observed in region-I, Fig. 3(a). Such behaviours suggest competition between the thermal effect and the magnetic interaction. The different behaviours of the \textit{a} and \textit{c}-parameters suggest magnetic anisotropy existing in the compound, Fig.3 (c). Interestingly, deviation from the paramagnetic behaviour is observed in region II around 260 K (T$_0$), inset to Fig. 1(b).

In the compound under study, there are two kinds of magnetic ions, namely Co and Nd. The magnetic interaction between the same kind of the ions within the layer is expected to occur through the intervening O ions through super exchange interactions. In literature, based on neutron diffraction results, it has been reported that there occurs FM interaction between the same kind of sublattice within the layer and FeM interaction between Co and Nd sublattices\citep{khalyavin2008}. In Fig.6 (e-f), we show the temperature dependent Co-O-Co bond angles between different Co ions. In region II, Co2-O8-Co3 bond angle increases by about 1.8$^{\circ}$ while the Co1-O7-Co3 bond angle decreases by about 0.9$^{\circ}$. As the increment in the former bond angle is more as compared to the latter, an increment in the \textit{a}-parameter is observed. Hence, the increment in the Co2-O8-Co3 bond angle in region-II is expected to lead to the increase in the hybridization between the 3\textit{d} states of the Co ions through the intervening O 2\textit{p} states. This leads to the onset of intralayer FM interaction between the Co ions having octahedral coordination at higher temperature as compared to the Co ions with square pyramidal coordination.

The observation of the decrement in the Nd-Nd bond distance within the layer suggests the onset of the FM interaction within the Nd sublattice occurring at lower temperature than the onset of the intralayer FM interaction within the Co sublattice.

In region III, the decrement in the \textit{a}-parameter below 120 K further suggests the strengthening of the FM interactions within both the sublattices. In the dc susceptibility, for the ZFC at 100 Oe (Fig. 1(b)), we observe a decrement below $T_1$ suggesting the strengthening of the interlayer interaction between Co and Nd sublattices.

\begin{figure}
\centering

\includegraphics [width = 0.45\textwidth]{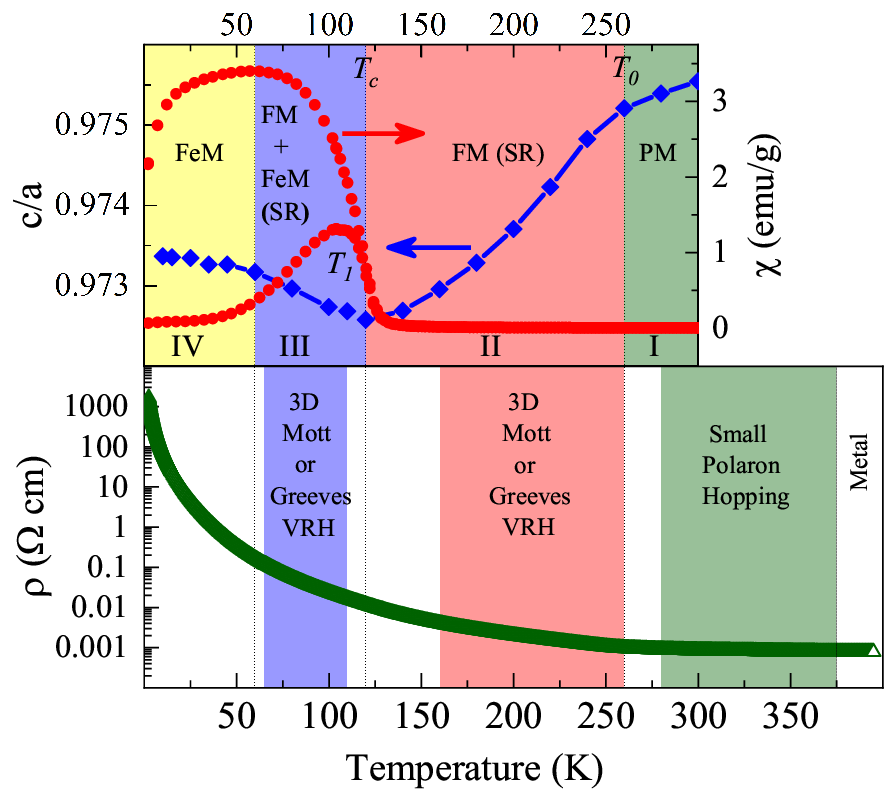}

\caption{ The top panel: Pictorial representation of the summary of the results of magnetic and structural studies. PM, FM, FeM, FM(SR), FeM (SR) represents parmagnetic, ferromagnetic, ferrimagnetic, short range FM and ferrimagnetic phases, respectively. 
The bottom panel: Summary of the results of transport studies.}

\end{figure}

The increment in the \textit{c}-parameter by $\sim$ 0.05\% covering regions III and IV suggests the competition between the intra and interlayer magnetic interaction in addition to the thermal effect. As mentioned above, in the Nd ion cage, the bending of the Co-O-Co bond angles occurs inward. In the temperature range 120 to 10 K, the Co2-O8-Co3 bond angle decreases by $\sim$1.8$^{\circ}$ while the Co1-O7-Co3 bond angle increases by $\sim$1.4$^{\circ}$. Corresponding to the two competing bond angles, there are two different kinds of Nd-O bonds. The shorter one is connected to the intervening O8 ions of the Co3 and Co2 octahedra and the longer ones to the intervening O7 ions connected to the Co3 octahedra and Co1 square pyramidal coordination.  The decrement in the Nd-O bonds lead to the increment in the hybridisation of the Nd 4\textit{f} states with the Co 3d states through Nd 4\textit{f} - Nd 5\textit{d} - O 2\textit{p} - Co 3\textit{d} pathway\citep{kundu2015interplay}. This leads to the onset of the interlayer FeM interaction between the Nd and Co sublattices. With further decrement in the temperature below 60 K, the increment in the \textit{c}-parameter is negligible suggesting the fact that the intralayer FM interactions in both the sublattices are completely established and hence to the setting up of the interlayer FeM interaction between the sublattices. It is important to note that from our results based on the Curie-Weiss fit, the values of $T_c$ obtained for Co and Nd sublattice are $\sim$ 120 K and $60\;$K, respectively. As there are two kinds of Nd-O bonds, it is expected that there will be distribution in the onset temperature of FeM interaction thereby leading to a broad peak around $T_1$. Such competing magnetic interactions in the regions III and IV could possibly lead to zero thermal expansion in the unit cell volume.

Hence, our structural results show signature of the onset of the short-range FM and FeM ordering below 260 and 120 K, respectively. It is interesting to note that around 260 K, we also observe change in the slope in the resistivity data, inset to Fig. 4(a). The observation of short-range ordering in region II and coexistence of short- and long-range magnetic orderings in region III could be the possible reason for the stabilization of variable range hopping behaviour in these regions.

In addition to the structural connection with the magnetism, it is important to note that the break in the inversion symmetry around the Co ions both in the octahedral and square pyramidal coordination, could lead to ferroelectric behaviour thereby indicating the possibility of ferroelectricity accompanied by magnetic effects existing in this compound. However, more experiments are required in this direction to confirm it. The pictorial representation of the structural, magnetic and transport studies are shown in Fig. 7. It is interesting to observe that in the heat capacity data, the magnetic transitions observed are quite broad suggesting the stabilization of magnetic short range ordering.

\subsection{DFT Results}

\begin{figure}

\includegraphics [width = 0.49\textwidth]{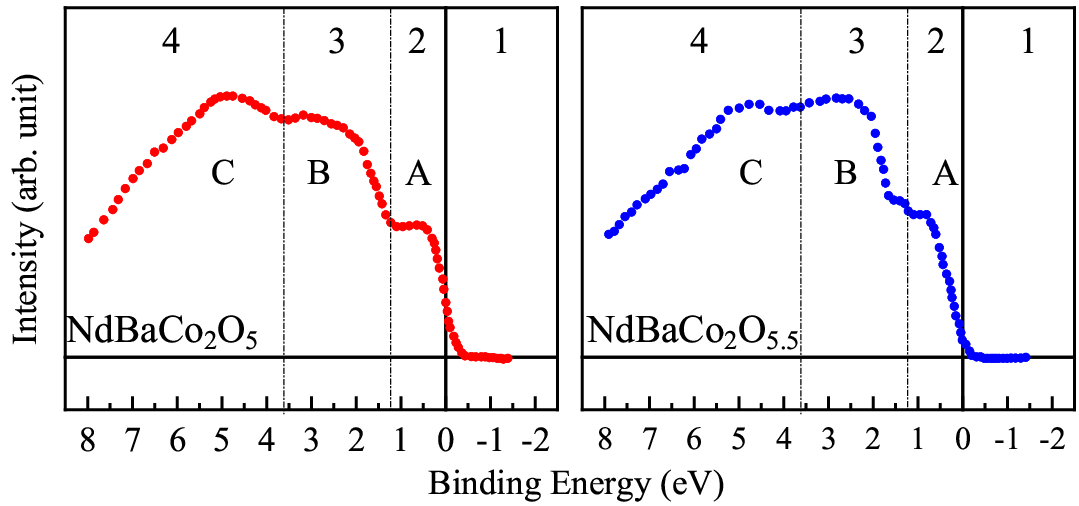}

\caption {Comparison of xps valance band spectra of NdBaCo$_2$O$_5$ and NdBaCo$_2$O$_{5.5}$.\citep{takubo2006}.}

\end{figure}

 In literature, to the best of our knowledge, on NdBaCo$_2$O$_{5+\delta}$ systems, only RT x-ray photoemission spectroscopic (xps) studies have been carried out on single crystalline compounds for  $\delta = 0, 0.5$\cite{takubo2006}. Keeping the reported experimental results in mind, to understand the electronic structure and effect of electron-electron correlation in these systems, spin polarised electronic structure calculations under DFT and DFT+\textit{U}  were performed for $\delta = 0.75$. Such study will also help in obtaining the value of the electron counts in the different orbitals and the value of magnetic moment.

 Based on the xps studies on NdBaCo$_2$O$_{5+\delta}$ for $\delta = 0, 0.5$ reported by Takubo \textit{et. al.}\citep{takubo2006}, the valence band spectra can be broadly divided into four regions. We mark the region 1 that covers the binding energy (BE) range -2.5 eV to $\varepsilon_f$; region 2 as $\varepsilon_f$ to 1.23 eV; regions 3 and 4 covering the range 1.23 eV to 3.62 eV and 3.62 eV to 8 eV, respectively. Here, we represent the features appearing in regions 2 to 4 as feature A, B and C, respectively, shown in Fig. 8. For the case of $\delta ={0}$ compound, the features A,B and C are observed around  0.84 eV, 2.5 eV and 4.64 eV, respectively. For $\delta ={0.5}$, it is around   0.64 eV, 2.5 eV and 4.65 eV, respectively. At the $\varepsilon_f$, the intensity is enhanced for the case of   $\delta ={0.5}$ as compared to  $\delta ={0}$. This behaviour is in line with the resistivity results where resistivity for $\delta ={0}$ is high compared to $\delta ={0.5}$\cite{taskin2006, sodaNBCO55,sodaNBCO5,solin2022}.

 In literature, features A, B and C are attributed to the major contribution from Co 3\textit{d}, O 2\textit{p} and Nd 4\textit{f} states, respectively. \cite{takubo2006,flavell2004resonant}

\begin{figure}

\includegraphics [width = 0.49\textwidth]{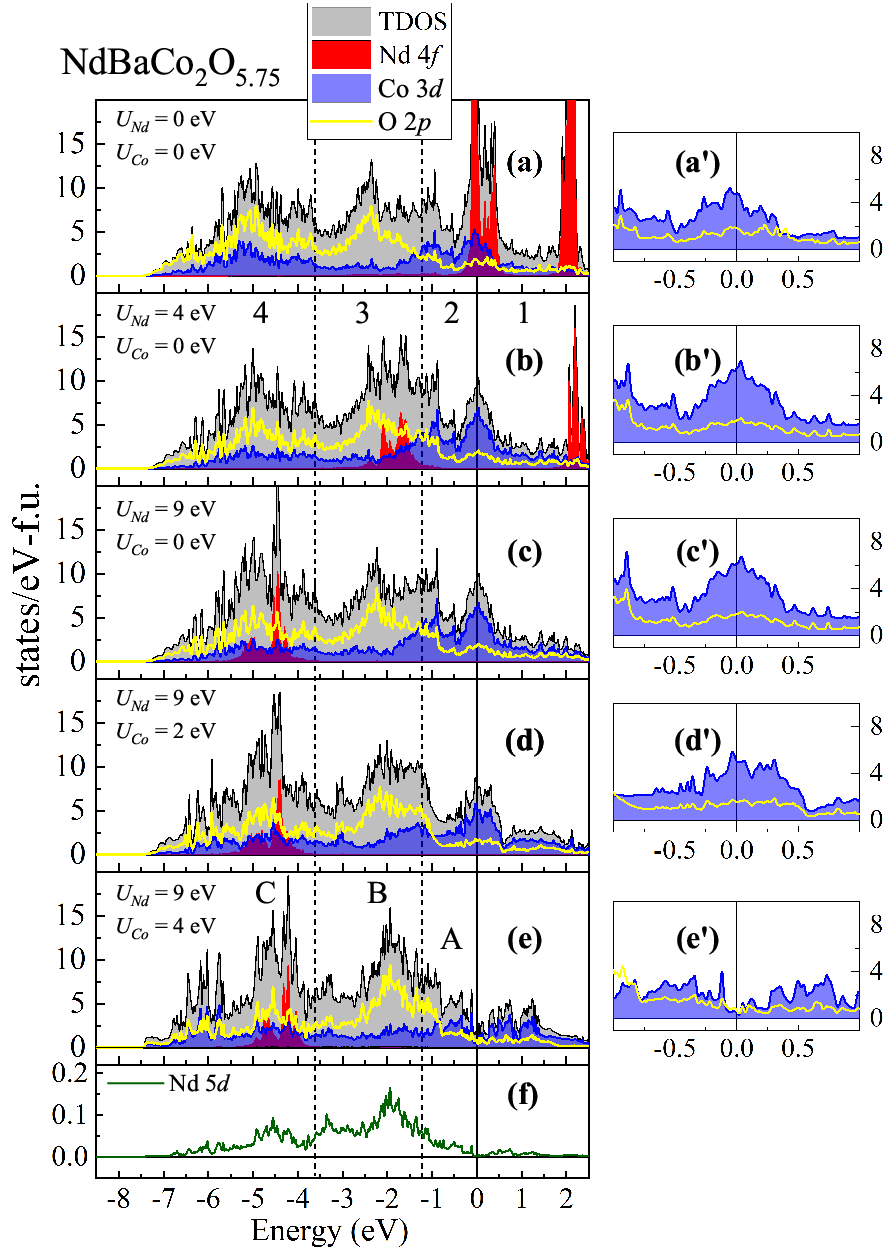}

\caption {DFT + \textit{U} calculation results on NdBaCo$_2$O$_{5.75}$, using structural parameters obtained from Reitveld refinement data for sample 1 at RT with different $U_{Nd}$ and $U_{Co}$ values, (a)-(e) TDOS and partial density of states (PDOS); (a')-(e') the corresponding PDOS near $\varepsilon_f$. and (f) PDOS for Nd 5\textit{d} from DFT + \textit{U} calculation using $U_{Nd}$ = 9 eV and $U_{Co}$ =4 eV.}

\end{figure}

\begin{table}
  
  \caption{Electron counts as obtained in the DFT+\textit{U} calculations using the value of $U_{Nd}$ = 9eV and $U_{Co}$ = 4eV, within the muffin tin sphere for different orbitals. }
  
  \begin{tabular}{c|c|c|c}

  & Electron Count &  & Electron Count\\
  \hline
  
  Nd  4\textit{f}& 3.000& O2 2\textit{p}&3.596\\
  6\textit{s}& 0.042  & 3\textit{s} & 0.084 \\
  5\textit{d}& 0.325  & 3\textit{d}& 0.016 \\
  6\textit{p}& 0.231 & 4\textit{f}& 0.003\\
  Ba 5\textit{p}& 4.491 & O3 2\textit{p}&3.606 \\
  6\textit{s}& 0.035 & 3\textit{s} & 0.084 \\
  4\textit{f}& 0.027 & 3\textit{d}& 0.014\\
  5\textit{d}& 0.116 & 4\textit{f}& 0.002\\
  Co1 3\textit{d}&6.134 & O5 2\textit{p}&3.624\\
  4\textit{s}& 0.114 & 3\textit{s} & 0.052\\
  4\textit{p}& 0.128 & 3\textit{d}& 0.016\\
  4\textit{f}& 0.012 & 4\textit{f}& 0.0032\\
  Co2 3\textit{d}&6.232 & O6 2\textit{p}&3.612\\
  4\textit{s}& 0.104 & 3\textit{s} & 0.052\\
  4\textit{p}& 0.124 & 3\textit{d}& 0.014\\
  4\textit{f}& 0.014 & 4\textit{f}& 0.004\\
  Co3 3\textit{d}& 6.136 & O7 2\textit{p}& 3.619\\
   4\textit{s}& 0.099 & 3\textit{s} & 0.0545\\
   4\textit{p}& 0.114 & 3\textit{d}& 0.012\\
  4\textit{f}& 0.0522 & 4\textit{f}& 0.0025 \\
  O1 2\textit{p}& 3.644& O8 2\textit{p}&3.635\\
  3\textit{s}& 0.096 & 3\textit{s} & 0.0605\\
  3\textit{d}& 0.016 & 3\textit{d}& 0.0125\\
  4\textit{f}& 0.0034 & 4\textit{f}& 0.0025\\

  \hline
  
  \end{tabular}
   
\end{table}

From the results of DFT calculations for NdBaCo$_2$O$_{5.75}$, we observe significant density of states (DOS) at $\varepsilon_f$, with dominant contribution from Nd 4\textit{f} states, shown in Fig. 9(a). This is not in line with the experimental spectra of such compounds, Fig.8. In the valence band xps studies, Takubo \textit{et. al.}\citep{takubo2006} have reported Nd 4\textit{f} states to be distributed from 2 to 8 eV BE. To match the position of Nd 4\textit{f} states as per the xps studies on NdBaCo$_2$O$_{5+\delta}$\citep{takubo2006} and also considering localised nature of 4\textit{f} states, DFT+\textit{U} calculations were performed by the inclusion of $U_{Nd}$ in Nd 4\textit{f} states. The value of $U_{Nd}$ was increased from 1 to 9 eV. With the increase in $U_{Nd}$, we observe splitting in the band associated with Nd 4\textit{f} states. This splitting in the band increases with increase in $U_{Nd}$, as observed in Fig. 9(a)\&(b). For  $U_{Nd}$ = 9 eV, we observe a decrement in the DOS at $\varepsilon_f$ with Nd 4\textit{f} states peaked around -4.5 eV. In the case of Nd metal, based on Nd 4\textit{d}-4\textit{f} resonant photoemission experiments\cite{gerken1985}, the Nd 4\textit{f} states are found to be peaked around 4.5 eV below $\varepsilon_f$. Based on this result, we have chosen the value for $U_{Nd}$ = 9 eV that gives a better representation of the electronic structure. It is important to note that based on resonant photoemission studies on (Gd/Dy)BaCo$_2$O$_{5.5}$ single crystals\citep{flavell2004resonant}, Gd and Dy 4\textit{f} states also lie around the similar BE positions as obtained in Gd and Dy metals, respectively. However, similar studies will be required to confirm the position of Nd 4\textit{f} states in the compound under study.

 To incorporate the localised nature of Co 3\textit{d} states in the electronic structure, in addition to $U_{Nd}$ = 9 eV, $U_{Co}$ = 1-4 eV (in steps of 1 eV) was added in the calculations. Fig. 9(d)\&(e) show the results for DFT+\textit{U} calculations using different values of $U_{Nd}$ and $U_{Co}$. For $U_{Nd}$ = 9 eV and $U_{Co}$ = 1 and 2 eV, total density of states (TDOS) around $\varepsilon_f$ remains almost the same. But with $U_{Co}$ = 3eV (not shown), a significant reduction in TDOS around $\varepsilon_f$ is observed. In literature, value of $U_{Co}$ $\sim$ 4 to 5 eV is generally used for cobalt based oxide systems\cite{wu2003RBCO5.5,wu2001TBCO5.5}. Considering this fact, calculations have been carried out until $U_{Nd}$ = 9eV and $U_{Co}$ = 4eV. However, comparison of the results of the calculations with xps or resonant photoemission studies of compound under study is vital to make any concluding remarks.

 In Table IV, the number of electrons within the muffin tin sphere for different states are shown. Based on the ionic model and for charge balance, the valence state of Nd, Ba and O is expected to be in +3, +2 and -2, respectively. Co is expected to stabilize in the mixed valence state of +3 and +4 in the ratio 3:1 in NdBaCo$_2$O$_{5.75}$\citep{khalyavin2008,lobanovsky2006}. It is important to note that the number of electrons for Nd 6\textit{s}, 6\textit{p}, 5\textit{d}; Ba 6\textit{s}, 5\textit{d}, 4\textit{f}; Co 4\textit{s}, 4\textit{p}, 4\textit{f} are expected to be zero, but finite number of electrons for these states are observed, Table IV. O 3\textit{s} orbital also has finite number of electrons, suggesting there is a intra transfer of electrons from O 2\textit{p} to O 3\textit{s}. Co 3\textit{d} states show the number of electrons more than expected. This suggests there is an inter transfer of electrons from O 2\textit{p} to  Co 3\textit{d}. From this, one can conclude that the valence state of the Co ions is less than +3.  It is important to note here that for the case $\delta$ = 0 compound, from the ionic model, the Co ions are expected to be stabilized in +3 and +2 oxidation states. From the xps studies, only signature of Co ion in +3 oxidation state has been observed\citep{takubo2006}.
 
 From the results of the calculations, the value of the spin moment for Nd ions is 2.97 $\mu_{B}$, indicating full spin polarisation. Value for spin moment for Co1, Co2 and Co3 is 2.05 $\mu_{B}$, 1.91 $\mu_{B}$ and 2.37 $\mu_{B}$, respectively. These values of spin moment combined with number of electrons in the different Cobalt ions indicate that all the Cobalt ions in the system stabilize in IS state. 
 
 Fig. 9(f) shows Nd 5\textit{d} states, having a similar distribution as of the O 2\textit{p} states near $\varepsilon_f$. This indicates hybridisation between O 2\textit{p} and Nd 5\textit{d} states. This suggests that magnetic interaction between Nd 4\textit{f}  and Co 3\textit{d} orbitals happens via Nd 4\textit{f} - Nd 5\textit{d} - O 2\textit{p} - Co 3\textit{d} route, as suggested in B site rare earth doped iron perovskites\cite{kundu2015interplay}.

\begin{figure}

\includegraphics [width = 0.49\textwidth]{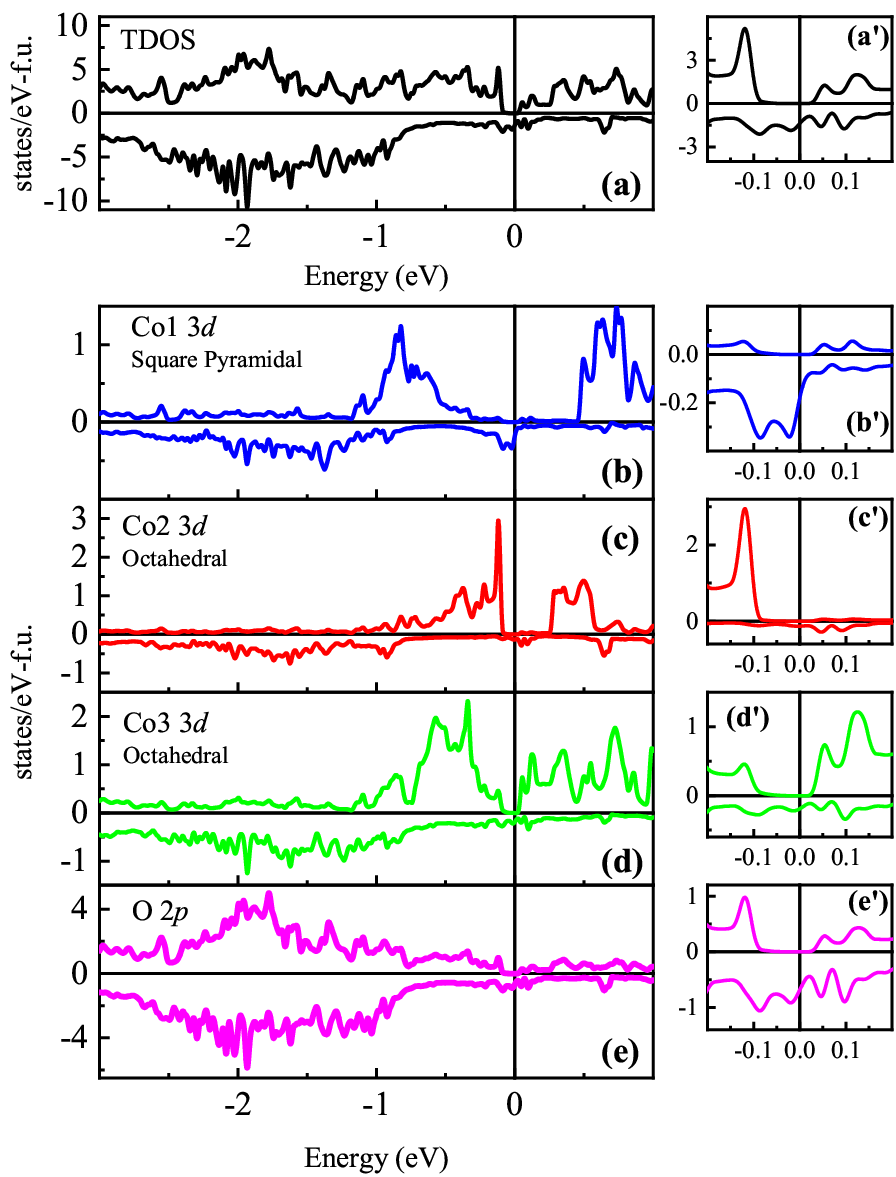}

\caption {DFT + \textit{U} calculation results on NdBaCo$_2$O$_{5.75}$, with $U_{Nd}$ = 9 eV and $U_{Co}$ = 4 eV values (a)-(e) spin polarised TDOS and PDOS; (a')-(e') spin polarised TDOS and PDOS.}

\end{figure}

To understand the significant reduction in TDOS around $\varepsilon_f$ with the inclusion of $U_{Nd}$ = 9 eV and $U_{Co}$ = 4 eV, spin polarised TDOS for the system was examined, Fig. 10(a). Our results show that at the $\varepsilon_f$, the TDOS for the up spin state develops a gap ($\sim$ 27 meV) while the down spin state shows finite DOS suggesting half metallicity. Such half metallic behaviour is observed only for above mentioned \textit{U} values. Fig. 10(b)-(e) shows the PDOS for different Co and O atoms. As the major contribution in the DOS at $\varepsilon_f$ is given by Co 3\textit{d} and O 2\textit{p} states, the existence of half metallicity in the system is again evident in the PDOS for Co 3\textit{d} and O 2\textit{p} states, Fig. 10(a')-(e').

 To understand the extent of exchange correlation in the Nd 4\textit{f} and Co 3\textit{d} states and to confirm the results of the calculations, it is important to perform the detailed photoemission spectroscopic (PES) studies. And also an insight into the spin state of the Co ions can be obtained by performing electronic structure calculations on low temperature magnetic structure existing in this compound. X-ray absorption spectroscopic studies as a function of temperature will also be helpful in unravelling the nature of spin state of the Co ions.

\section{Summary}

Polycrystalline samples of oxygen-deficient layered double perovskite cobaltite, NdBaCo$_2$O$_{5+\delta}$ ($\delta$ = 0.65) are studied using x-ray diffraction, dc susceptibility, resistivity and heat capacity measurements, combined with electronic structure calculations using DFT+\textit{U} method. Our results show that the compound crystallises in tetragonal structure (\textit{P4/mmm}) with 222 superstructure and displays various phase transitions namely paramagnetic to ferromagnetic ($\sim$ 120 K) and then to ferrimagnetic phase ($\sim$ 60 K). The structural analysis reveals signature of the stabilisation of short-range magnetic interactions well above the temperatures of its stabilisation at the long-range level. In the regions of stablisation of short range ordering, we observe Griffiths phase (120 K $<$ T $<$ 260 K) in the magnetic studies and   VRH in the transport studies. The zero thermal expansion in volume observed at low temperature appears to be due to the competition between magnetic interactions and also could explain the origin of the negative thermal expansion of the \textit{c}-parameter. Our electronic structure calculations show importance of electron-electron correlation in defining the electronic structure for this system. Calculations show that Co ions stabilize in an IS state, having oxidation state less than +3. Half metallicity in the system is also observed from the results of the calculations. Our results suggest connection between magnetism and ferroelectricity in this compound. We believe that our results on the valence state of the Co ion and the link between the structure and magnetism will be helpful in dispelling the confusion especially with regard to the nature of magnetism existing on such double perovskite systems. This would further help in paving the way for the manifestation of new applications.

\bibliographystyle{apsrev4-1}
\bibliography{Manuscript}

\end{document}